\documentclass{aa}  
\usepackage{color}
\usepackage{natbib}
\usepackage{graphicx}
\usepackage{txfonts}
\usepackage{comment}
\usepackage[linkcolor=blue]{hyperref}
\newcommand{\ha}{\hbox{H$\alpha$}}
\newcommand{\hb}{\hbox{H$\beta$}}
\newcommand{\pa}{\hbox{Pa$\alpha$}}
\newcommand{\pb}{\hbox{Pa$\beta$}}
\newcommand{\hei}{\ion{He}{i}} 
\newcommand{\heii}{\ion{He}{ii}}

\begin{document}

   \title{Novel calibrations of	virial black hole mass estimators in active galaxies based on X--ray luminosity and optical/NIR emission lines}

   \author{F. Ricci
          \inst{1}
          \and
          F. La Franca\inst{1} 
         \and
         F. Onori\inst{2,3} 
         \and
         S. Bianchi\inst{1} 
          }

   \institute{Dipartimento di Matematica e Fisica, Universit\`a Roma Tre, via della Vasca Navale 84, 00146 Roma, Italy\\
              \email{riccif@fis.uniroma3.it}
         \and
             SRON, Netherlands Institute for Space
             Research, Sorbonnelaan 2, 3584 CA, Utrecht, the Netherlands\\
         \and
             Department of Astrophysics/IMAPP, Radboud University, P.O. Box9010, 6500 GL Nijmegen, the Netherlands\\    
             }

   \date{Received July 25, 2016; accepted October 4, 2016}
\titlerunning{Novel virial BH masses calibrations}
\authorrunning{F. Ricci et al.}
 
  \abstract
   {Accurately weigh the masses of super massive black holes (BHs) in active galactic nuclei (AGN)
   	is currently possible for only a small group of local and bright broad-line AGN 
   	through reverberation mapping (RM).
    Statistical demographic studies can be carried out 
   	considering the empirical scaling relation between the size of the broad line region (BLR) and the 
   	AGN optical continuum luminosity.
	However,
   	there are still biases against low-luminosity or reddened AGN,
   	in which the rest-frame optical radiation can be severely absorbed or diluted by the host galaxy 
   	and the BLR emission lines could be hard to detect. }
   {Our purpose is to widen the applicability of virial-based single-epoch (SE) relations to 
   	reliably measure the BH masses also for low-luminosity or intermediate/type 2 AGN that are missed by 
   	current methodology. We achieve this goal by calibrating virial relations 
   	based on unbiased quantities: 
   	the hard X--ray luminosities, in the 2-10 keV and 14-195 keV bands, that are 
   	less sensitive to galaxy contamination,
   	and the full-width at half-maximum (FWHM) of the most important rest-frame near-infrared (NIR) and optical BLR emission lines. }
   { We built a sample of RM AGN having both X--ray luminosity and broad optical/NIR FWHM measurements available in order to calibrate new virial BH mass estimators.}
   { We found that the FWHM of the \ha, \hb\ and NIR lines (i.e. \pa, \pb\ and \hei\ $\lambda$10830) all correlate 
   	each other having negligible or small offsets. 
   	This result allowed us to derive virial BH mass estimators 
   	based on either 
   	the 2-10 keV or 14-195 keV luminosity. We took also into account the recent determination of the 
   	different virial
   	coefficients $f$ for pseudo and classical bulges.
   	By splitting the sample according to the bulge type 
   	and adopting separate $f$ factors \citep{hk14}
   	we found that 
   	our virial relations predict 
   	 BH masses 
   	of AGN hosted in pseudobulges 
   	$\sim$0.5 dex smaller than in classical bulges. 
   	Assuming the same average $f$ factor for both population, a difference of $\sim$0.2 dex 
   		is still found.
   	}
   {}


   \maketitle
%

\section{Introduction}
Supermassive black holes (SMBHs; having black hole masses $M_{\rm{BH}}=10^5 - 10^9 \, {\rm M_\odot}$) 
are observed to be common, 
hosted in the central spheroid in the majority of local galaxies.  
This discovery, combined with the observation 
of striking 
empirical relations between BH mass and host galaxy properties, 
opened in the last two decades an exciting era in extragalactic astronomy. 
In particular the realization that 
BH mass correlates strongly 
with the stellar luminosity, mass 
and velocity dispersion of the bulge 
\citep[for a review]{dressler89, kormendyrichstone95, magorrian98, ferrarese00, gebhardt00, marconihunt03, sani11, graham16}
suggests that  
SMBHs may play a crucial role in regulating many aspects of galaxy formation and evolution 
\citep[e.g. through AGN feedback,][]{silkrees98, fabian99, dimatteo05, croton06, sijacki07, ostriker10, fabian12, king14}.

One of the most reliable and direct ways to measure the mass 
of a SMBH 
residing in the nucleus of 
an active galaxy (i.e. an active galactic nucleus, AGN) is reverberation mapping 
\citep[RM,][]{blandformcknee82, peterson93}.
The RM technique takes advantage of AGN
flux variability to constrain black hole masses through time-resolved
observations.
With this method, the distance $R=c\tau$ of the Broad Line Region (BLR)
is estimated by 
measuring the time lag $\tau$ of the response of a permitted broad  
emission line to variation of the 
photoionizing primary continuum emission.
Under the hypothesis of a virialized BLR, 
whose dynamics are gravitationally dominated by the central SMBH,
$M_{\rm{BH}}$ is simply related to the velocity of the emitting gas clouds, $\Delta {\rm v}$, 
and to the size $R$ of the BLR, i.e. $M_{\rm{BH}} = \Delta {\rm v}^2 R G^{-1}$ ,
where G is the gravitational constant.
Usually the width $\Delta W$  of a doppler-broadened emission line 
(i.e. the full-width at half-maximum, FWHM, or the line dispersion, $\sigma_{\rm{line}}$) 
is used as a proxy of the real gas velocity $\Delta {\rm v}$, after introducing a 
virial factor $f$ which takes into account our ignorance on the structure, geometry and kinematics of the BLR 
\citep{ho99, wandel99, kaspi00}
\begin{equation}
M_{\rm{BH}} = f \frac{\Delta W^2 R}{G}\, .
\end{equation}
Operatively, the RM BH mass is equal to $f\times M_{\rm{vir}}$, where the virial mass $M_{\rm{vir}}$
is $\Delta W^2 R G^{-1}$.
In the last decade, the $f$ factor 
	has been studied by 
	several authors, finding values in the range 2.8 -- 5.5 
	\citep[if the line dispersion $\sigma_{\rm{line}}$ is used, see e.g.][]{onken04, woo10, graham11, park12, grier13}.
This quantity is statistically determined 
by normalizing
the RM AGN to the relation between BH mass and bulge
stellar velocity dispersion 
\citep[$M_{\rm{BH}} - \sigma_\star$ relation; see][]{ferrarese02, tremaine02, hu08,
		gultekin09, grahamscott13,
		mcconnellma13, kormendyho13, savorgnan15, sabra15}
observed in  
local inactive galaxies with direct BH mass measurements.
However, recently \citet{shankar16} claimed that 
	the previously computed $f$ factors could have been artificially 
	increased by a factor of at least $\sim$3 because of a presence 
	of a selection bias in the calibrating samples, in favour of the more massive 
	BHs. 
\citet{kormendyho13} significantly updated the
$M_{\rm{BH}} - \sigma_\star$ relation for inactive galaxies, 
highlighting a large and systematic difference between the relations
for pseudo and classical bulges/ellipticals.
It should be noted that the classification 
	of galaxies into 
	classical and pseudo bulges is a difficult task\footnote{Some authors 
			have also discussed how could be neither appropriate nor possible 
			to reliably separate bulges into one class or another \citep{graham14}, 
			and that in some galaxies there is evidence of 
			coexistence of classical bulges and pseudobulges \citep{erwin15,dullo16}.}, 
	which depends on a number of selection criteria, 
	which should not be used individually
	\citep[e.g. not only the 
	Sersic index \citep{sersic68} $n < 2$ condition to classify 
	a source as a pseudo bulge; see][]{kormendyho13, kormendy16}.

The results on the $M_{\rm{BH}} - \sigma_\star$ found by \citet{kormendyho13} prompted \citet{hk14} to calibrate the $f$ factor separately for the two bulge populations
\citep[for a similar approach see also][that derived different 
$M_{\rm{BH}} - \sigma_\star$ relations and $f$ factors
for barred and non-barred galaxies]{graham11}, getting $f_{\rm{CB}}=6.3\pm1.5$ for elliptical/classical and $f_{\rm{PB}}=3.2\pm0.7$ for pseudo bulges when the \hb\
$\sigma_{\rm{line}}$ (not the 
FWHM)\footnote{Note that if the FWHM instead of the $\sigma_{\rm{line}}$
		is used, the virial coefficient $f$ has to be properly scaled 
		depending on the FWHM/$\sigma_{\rm{line}}$ ratio \citep[see e.g.][for details]{onken04, collin06} .} is used to compute the virial mass.

However, RM campaigns are time-consuming and are accessible 
only for a handful of nearby (i.e. $z\lesssim$0.1) AGN. The finding of a tight relation between the distance of the BLR clouds $R$ and 
the AGN continuum luminosity $L$ 
\citep[$R \propto L^{0.5}$,][]{bentz06, bentz13}, has allowed to calibrate new single-epoch (SE) relations that can be used on larger samples of AGN, such as

\begin{equation}\label{eq:cal}
\log \left(\frac{M_{ \mathrm{BH}} }{M_\odot}\right) = 
a + b \log \left[ \left(\frac{L}{10^{42}~ \mathrm{erg \, s^{-1}} }\right)^{0.5}  \left(\frac{FWHM}{10^4 ~\mathrm{km\, s^{-1} }}\right)^2 \right]  
\, ;
\end{equation}
where the term $\log( L^{0.5} \times FWHM^2 )$ is generally known as virial product (VP). 
These SE relations have a typical spread of $\sim$ 0.5 dex 
\citep[e.g.][]{mj02,vp06}
and are calibrated using either 
the broad emission line or
the continuum luminosity
\citep[e.g. in the ultraviolet and optical, mostly at 5100 \AA, $L_{5100}$; see the review by][and references therein]{shen13}
 and the FWHM (or the $\sigma_{\rm{line}}$) of optical emission lines\footnote{As 
 the calibrating RM masses are computed by measuring the 
 BLR line width and its average distance $R=c\tau$, the fit of Equation \ref{eq:cal}
 corresponds, strictly speaking, to 
 the fit of the $\tau$ versus $L$ relation \citep[e.g.][]{bentz06}. }, 
 such as \hb, \ion{Mg}{ii} $\lambda$2798, \ion{C}{iv} $\lambda$1549 
\citep[even though the latter is still debated; e.g.][]{bl05, sl12, denney12, runnoe13}.

These empirical scaling relations have some problems though:
\begin{itemize}
\item Broad \ion{Fe}{ii} 
emission in type 1 AGN (AGN1) can
add ambiguity in the determination of the optical continuum luminosity at 5100 \AA. 
\item In low-luminosity AGN, host galaxy starlight dilution 
can severely affect the AGN ultraviolet and optical continuum emission.
Therefore in such sources
it becomes very challenging, if not impossible at all, 
to isolate the AGN contribution unambiguously.
\item The \hb\ transition is at least a factor of three weaker
than \ha\, and so from considerations of signal-to-noise ratio (S/N)
alone, \ha, if available, is superior to \hb. In practice, in some
cases \ha\ may be the only line with a detectable broad component
in the optical \citep[such objects are known as Seyfert 1.9 galaxies;][]{osterbrock81}.
\item The optical SE scaling relations are completely biased against
type 2 AGN (AGN2) which lack broad emission lines in the rest-frame optical spectra. 
However, several studies have shown that most AGN2 exhibit faint components
of broad lines if observed with high ($\gtrsim$20) S/N in the rest-frame near-infrared (NIR), 
where the dust absorption is less severe than in
the optical 
\citep{veilleux97,riffel06, cai10, onori16inpress}.
Moreover, 
some studies have shown that NIR lines (i.e. \pa\ and \pb) can be reliably used 
to estimate the BH masses in AGN1 
\citep{kim10, kim15,landt13}
and also for intermediate/type 2 AGN \citep{LF15,LF16}.
\end{itemize}

In an effort to 
widen the applicability of this kind of relations to classes of AGN that
would otherwise be inaccessible using the conventional methodology 
(e.g. galaxy-dominated low-luminosity sources,
type 1.9 Seyfert and type 2 AGN), \citet{LF15} fitted new 
virial BH mass estimators based on intrinsic (i.e. absorption corrected)
hard, 14-195 keV, X--ray luminosity, which is thought to be produced by the hot corona 
via Compton scattering of the ultraviolet and optical photons coming from the 
accretion disk \citep{hm91, hmg94, hmg97}.
Actually the X--ray luminosity $L_{\rm{X}}$ is known to be empirically related to 
the dimension of the BLR, as it is observed for the optical 
continuum luminosity of the AGN accretion disk \citep[e.g.][]{maiolino07, greene10}.
Thanks to these $R-L_{{\rm X}}$ empirical scaling relations, 
also \citet{bongiorno14} have derived virial relations based on the \ha\ width and
on the hard, 2-10 keV, X--ray luminosity, that
is less affected by galaxy obscuration (excluding severely absorbed, Compton Thick, AGN: $N_{\rm{H}}>10^{24}$ cm$^{-2}$). 
Indeed in the 14-195 keV band up to $N_{\rm{H}}<10^{24}$ cm$^{-2}$ the absorption is negligible while in the 
2-10 keV band the intrinsic X--ray luminosity can be recovered after measuring the $N_{\rm{H}}$ column density via X--ray spectral fitting. 

Recently \citet{hk15} showed that the BH masses of RM AGN 
	correlates tightly and linearly with the optical VP (i.e. FHWM(\hb)$^2 \times L_{5100}^{0.5}$) 
	with different logarithmic zero points for elliptical/classical and pseudo bulges. 
	They used the updated database of RM AGN with bulge classification from \citet{hk14} 
	and adopted the virial factors separately for classical ($f_{\rm{CB}}=6.3$) and pseudobulges ($f_{\rm{PB}}=3.2$).	

Prompted by these results, in this paper we present an update of the calibrations 
of the virial relations based on the hard 14-195 keV X--ray luminosity published in \citet{LF15}. 
We extend these calibrations to the 2-10 keV X--ray luminosity and to the most intense optical and NIR emission lines, i.e. 
\hb\ $\lambda$4862.7 \AA, \ha\ $\lambda$6564.6 \AA, \hei\ $\lambda$10830.0 \AA, \pb\ $\lambda$12821.6 \AA\ and \pa\ $\lambda$18756.1 \AA.
In order to minimize the statistical uncertainties 
in the estimate of the parameters $a$ and $b$ of the virial 
relation (Equation \ref{eq:cal}), we have verified that 
reliable statistical correlations do exist among 
the hard X--ray luminosities, $L_{2-10 \, \rm{keV}}$ and $L_{14-195 \, \rm{keV}}$,
and among the optical and NIR emission lines \citep[as already found by other studies:][]{gh05, landt08, mejia16}.
These correlations allowed us 1) to 
compute, using the total dataset, average FWHM and $L_X$ 
for each object and then derive more statistically robust 
virial relations and 2) to compute our BH mass estimator 
using any combination of $L_X$ and 
optical or NIR emission line width.

We will proceed as follows: 
Sect. \ref{sec:data} presents the RM AGN dataset; 
in Sect. \ref{sec:lines} 
we will test whether 
the optical (i.e. \ha\ and \hb) and NIR (i.e. \pa, \pb\ and \hei\ $\lambda$10830 \AA, hereafter \hei) 
emission lines probe similar region in the BLR; 
in Sect. \ref{sec:cal} new calibrations of the
virial relations based on the average hard X--ray luminosity and 
the average optical/NIR emission lines width, 
taking (or not) into account the bulge classification, are presented; finally Sect. \ref{sec:concl} addresses the discussion of our findings and the conclusions.
Throughout the paper we assume a flat $\Lambda$CDM cosmology with cosmological parameters: $\Omega_\Lambda$ = 0.7, $\Omega_{\rm{M}}$ = 0.3 and $H_0$ = 70 km s$^{-1}$ Mpc$^{-1}$. Unless otherwise stated, all the quoted uncertainties are at 68\% (1$\sigma$) confidence level.

\begin{sidewaystable*}
	\caption{Properties of the RM AGN with both bulge classification and hard X--ray luminosity. }             
	\label{tab:1}
	\centering                          
	\begin{tabular}{l c c c c c c c c c r c}
		\hline\hline 
		\noalign{\smallskip}
		Galaxy    &$\log L_{2-10{\rm \, keV}}$	&$\log L_{14-195{\rm \, keV}}$&FWHM \hb\ &FWHM \ha\	&FWHM \pa\	&FWHM \pb\ 	&FWHM \hei\	&ref &$M_{\rm{vir}}$ &ref &Bulge \\
		&[erg s$^{-1}$]&[erg s$^{-1}$] &[km s$^{-1}$]&[km s$^{-1}$]	&[km s$^{-1}$]		&[km s$^{-1}$]	&[km s$^{-1}$]	&OPT/NIR&[$10^6 M_\odot$]	&$M_{\rm{vir}}$ 	&type \\
		(1) & (2) & (3) & (4) & (5) & (6) & (7) & (8) & (9) & (10) & (11) & (12) \\
		
		\noalign{\smallskip}
		\hline
		
		3C 120  		&44.06 &44.38	&1430 &2168	 &\dots		&2727		&\dots		&G12,K14/L13	&$12.2^{+0.9}_{-0.9}$		&G13	&CB \\
		3C 273 			&45.80 &46.48	&3943 &2773	 &2946		&2895		&3175		&L08,K00/L08	&$161^{+34}_{-34}$			&P04  	&CB \\
		Ark 120			&43.96 &44.23	&5927 &4801	 &5085		&5102 		&4488		&L08			&$23.4^{+4.0}_{-5.7}$		&G13	&CB \\
		Arp 151			&\dots &43.30	&3098 &1852	 &\dots		&\dots		&\dots		&B09,B10		&$1.1^{+0.2}_{-0.1}$		&G13	&CB \\
		Fairall 9 		&43.78 &44.41 	&6000 &\dots &\dots		&\dots 		&\dots		&P04	&$54.3^{+11.8}_{-11.6}$				&HK14 	&CB \\
		Mrk 79			&43.11 &43.72	&3679 &3921	 &3401		&3506 		&2480\tablefootmark{\textdagger}		&L08			&$19.2^{+4.5}_{-7.4}$		&G13	&CB \\
		Mrk 110			&43.91 &44.22	&2282 &1954	 &1827		&1886 		&1961		&L08			&$5.2^{+1.3}_{-2.1}$		&G13	&CB \\
		Mrk 279     	&\dots &43.92	&5354 &\dots &\dots		&3546		&\dots 		&P04/L13		&$7.2^{+1.1}_{-1.1}$		&G13	&PB \\
		Mrk 290			&43.08 &43.67	&5066 &4261	 &3542		&4228 		&3081		&L08			&$3.9^{+0.4}_{-0.3}$		&HK14	&CB \\
		Mrk 335			&43.44 &43.45	&2424 &1818	 &1642		&1825 		&1986		&L08			&$2.98^{+0.63}_{-0.68}$		&HK14	&CB \\
		Mrk 509     	&44.02 &44.42	&3947 &3242	 &3068		&3057 		&2959		&L08			&$22.2^{+1.0}_{-1.0}$		&G13	&CB \\
		Mrk 590 		&43.04 &43.42	&9874\tablefootmark{\textdagger}&4397		&4727 		&3949		&3369		&L08			&$7.3^{+1.2}_{-1.6}$		&G13    &PB \\
		Mrk 771	 		&43.47 &44.11	&3828 &2924	 &\dots		&\dots		&\dots		&P04,K00		&$16.0^{+2.7}_{-2.6}$		&G13	&PB \\
		Mrk 817			&43.46 &43.77	&6732 &5002	 &4665		&5519 		&4255		&L08			&$14.6^{+2.2}_{-2.5}$		&G13	&PB \\
		Mrk 876			&44.23 &44.73	&8361 &\dots &5505		&6010		&5629		&L08			&$50.8^{+21.3}_{-21.5}$		&P04   	&CB \\
		Mrk 1310		&41.34 &42.98	&2409 &561\tablefootmark{\textdagger}	    &\dots		&\dots		&\dots		&B09,B10		&$0.47^{+0.20}_{-0.17}$		&G13	&CB \\
		Mrk 1383		&44.18 &44.52	&7113 &5430	 &\dots		&\dots		&\dots		&P04,K00		&$373.3^{+68.7}_{-71.3}$	&G13	&CB \\
		Mrk 1513 		&43.56 &\dots	&1781 &\dots &1862 		&\dots		&\dots		&G12/L13		&$22.7^{+3.4}_{-3.4}$		&HK14	&PB \\
		NGC 3227    	&41.57 &42.56	&3939 &3414	 &\dots		&2934 		&3007		&L08			&$5.2^{+2.0}_{-2.1}$		&G13	&PB \\
		NGC 3516		&42.46 &43.31	&\dots &\dots&\dots		&4451 		&\dots		&	L13     	&$7.2^{+0.7}_{-0.6}$		&G13	&PB \\
		NGC 3783	   	&43.08 &43.58	&5549 &5290	 &\dots		&3500		&5098		&On+			&$4.4^{+0.7}_{-0.8}$		&G13	&PB \\
		NGC 4051 		&41.44 &41.67	&\dots &\dots&\dots		&1633 		&\dots		&	L13			&$0.5^{+0.1}_{-0.1}$		&G13	&PB \\
		NGC 4151 		&42.53 &43.12	&4859 &5248	 &\dots		&4654 		&2945\tablefootmark{\textdagger}		&L08			&$8.4^{+0.9}_{-0.5}$		&G13	&CB \\
		NGC 4253		&42.93 &42.91	&1609 &1013	 &\dots		&\dots		&\dots		&B09,B10		&$0.3^{+0.2}_{-0.2}$		&G13	&PB \\
		NGC 4593 		&42.87 &43.20	&4341 &3723	 &\dots		&3775 		&3232		&L08			&$2.1^{+0.4}_{-0.3}$		&G13	&PB \\
		NGC 4748		&42.63 &42.82	&1947 &1967	 &\dots		&\dots		&\dots		&B09,B10		&$0.7^{+0.2}_{-0.2}$		&G13	&PB \\
		NGC 5548		&43.42 &43.72	&10000\tablefootmark{\textdagger}&5841	&4555		&6516 		&6074		&L08			&$13.8^{+1.7}_{-2.0}$		&G13	&CB \\
		NGC 6814	 	&42.14 &42.67	&3323 &2909	 &\dots		&\dots		&\dots		&B09,B10		&$3.7^{+0.5}_{-0.5}$		&G13	&PB \\
		NGC 7469     	&43.23 &43.60	&1952 &2436	 &450\tablefootmark{\textdagger}		&1758 		&1972		&L08			&$4.8^{+1.4}_{-1.4}$		&G13	&PB \\
		PG 0026+129 	&\dots &44.83	&2544 &1457	 &1748		&\dots		&\dots		&P04,K00/L13	&$56.8^{+13.6}_{-12.9}$		&HK14	&CB \\
		PG 0052+251 	&44.64 &44.95	&5008 &2651	 &4114 		&\dots		&\dots		&P04,K00/L13	&$92.6^{+6.8}_{-6.4}$		&HK14	&CB \\
		PG 0804+761 	&44.44 &44.57	&3053 &2719	 &2269 		&\dots		&\dots		&P04,K00/L13	&$88.5^{+6.9}_{-8.7}$		&HK14	&CB \\
		PG 0844+349 	&43.70 &\dots 	&2288 &1991	 &2183 		&2377  		&2101 		&L08			&$5.0^{+2.2}_{-0.7}$		&HK14	&CB \\
		PG 0953+414 	&44.69 &\dots 	&3071 &\dots &\dots 	&\dots 		&\dots 		&P04 			&$54.0^{+8.5}_{-8.5}$ 		&HK14 	&CB \\
		PG 1211+143		&43.73 &\dots 	&2012 &1407	 &1550 		&\dots		&\dots		&P04,K00/L13	&$16.7^{+2.2}_{-6.2}$		&HK14	&CB \\
		PG 1307+085		&44.08 &\dots 	&5059 &3662	 &2955 		&\dots		&\dots		&P04,K00/L13	&$122^{+12.1}_{-11.8}$		&HK14	&CB \\
		PG 1411+442 	&43.40&\dots 	&2801 &2123	 &\dots		&\dots		&\dots		&P04,K00		&$26.9^{+7.7}_{-4.8}$		&HK14	&CB \\

		\noalign{\smallskip}
		\hline

	\end{tabular}
	
	\tablefoot{Columns are: (1) galaxy name; (2) and (3) logarithm of the 2-10 keV and 14-195 keV band luminosities; 
		(4) to (8) are the FWHMs of \hb, \ha, \pa, \pb\ and \hei; (9) references for the optical and 
		NIR emission lines, where: G12 is \citet{grier12}, K14 is \citet{kollatschny14}, L13 is \citet{landt13}, L08 is \citet{landt08}, K00 is \citet{kaspi00}, B09 is \citet{bentz09}, B10 is \citet{bentz10}, P04 is \citet{peterson04}, On+ is \citet{onori16inpress}; (10) virial BH masses (to be multiplied for the virial $f$ factor in order to obtain the 
			black hole mass $M_{\rm{BH}}$); (11) references for the virial masses, where:
		G13 is \citet{grier13}, HK14 is \citet{hk14} and the other labels are the same used in the references of column (9); 
		(12) bulge classification of each galaxy, CB = classical bulge or elliptical; PB = pseudobulge.
			From compilation of \citet{hk14}, which also contains references to original data sources.\\
		\tablefoottext{\textdagger}{Measurements considered outliers, see Sect. \ref{sec:lines} for more details.}}

\end{sidewaystable*}

\begin{table*}
	\caption{Properties of the additional AGN1 database used in the emission line relation analysis. }             
	\label{tab:1b}
	\centering                          
	\begin{tabular}{l c c c c c c}
		\hline\hline 
		\noalign{\smallskip}
		Galaxy    & FWHM \hb\ &FWHM \ha\	&FWHM \pa\	&FWHM \pb\ 	&FWHM \hei\	&ref \\
		          & [km s$^{-1}$] &[km s$^{-1}$]	&[km s$^{-1}$]		&[km s$^{-1}$]	&[km s$^{-1}$]	& OPT/NIR \\
		(1) & (2) & (3) & (4) & (5) & (6) & (7) \\
		
		\noalign{\smallskip}
		\hline
		
H 1821+643 	&	6615	&	5051	&	\dots &	5216  &	4844 	&L08	\\	
H 1934-063 	&	1683	&	1482	&	1354  & 1384  &	1473 	&L08	\\	
H 2106-099 	&	2890	&	2368	&	1723  & 2389  &	2553 	&L08	\\	
HE 1228+013 	&	2152	&	1857	&	1916  & 1923  &	1770 	&L08	\\	
IRAS 1750+508  &  	2551	&  	2323	&	\dots &	1952  &	1709	&L08	\\	
Mrk 877 		&	6641	&	4245	&	\dots &	\dots &	\dots	&K00,P04\\	
PDS 456		&	3159	&	\dots	&	2022  & 2068  &	\dots	&L08	\\	

SBS 1116+583A	&3668 &2059	 &\dots		&\dots		&\dots		&B09,B10	 \\

		\noalign{\smallskip}
		\hline

	\end{tabular}
	
	\tablefoot{Columns are: (1) galaxy name; 
		(2) to (6) are the FWHMs of \hb, \ha, \pa, \pb\ and \hei; (7) references for the optical and 
		NIR emission lines, where: L08 is \citet{landt08}, K00 is \citet{kaspi00}, P04 is \citet{peterson04},
		B09 is \citet{bentz09} and B10 is \citet{bentz10}.}
	
\end{table*}

\section{Data}\label{sec:data}
As we are interested in expanding the applicability of SE relations, we decided to 
use emission lines that can be more easily measured also in low-luminosity or obscured sources, such as 
the \ha\ or the \pa, \pb\ and \hei. 
Moreover, we want also to demonstrate that such lines can give reliable estimates as the \hb\ \citep[see e.g.][]{gh05, landt08, kim10, mejia16}.
For this reason, we built our sample starting from the database of 
\citet{hk14}
that lists 
43 RM AGN (i.e. $\sim$90\% of all 
	the RM black hole masses available in the literature), all having bulge type classifications
based on the  
criteria of 
\citet[Supplemental Material]{kormendyho13}. In particular, 
\citet{hk14} used the most common condition to classify as classical bulges those galaxies having
Sersic index \citep{sersic68} $n < 2$. However when the nucleus is too bright 
this condition is not totally reliable due to the difficulty of 
carefully measure the bulge properties. In this case \citet{hk14} adopted the condition 
that the bulge-to-total light fraction should be $\lesssim 0.2$ 
\citep[e.g.][but see \citealt{gw08} for a
discussion on the uncertainties of this selection criterion]{fisherdrory08,gadotti09}.
In some cases, additional clues came from the detection of 
circumnuclear rings and other signatures of ongoing central star formation.\\
It should be noted that an offset ($\sim$0.3 dex) 
is	observed in the $M_{\rm{BH}} -\sigma_\star$ diagram 
both when the galaxies are divided into barred and unbarred 
\citep[e.g.][]{graham08, graham11, grahamscott13} and 
into classical and pseudobulges \citep{hu08}.
However, the issue on the bulge type classification and 
on which host properties better discriminate the $M_{\rm{BH}} -\sigma_\star$ relation 
is beyond the scope of this work, and in the following we will adopt the bulge type
classification as described in \citet{hk14}.

Among this sample, we selected those AGN having hard 
X--ray luminosity and at least one emission line width available among \hb, \ha, \pa, \pb\ and \hei.
The data of 3C 390.3 were excluded since it clearly shows a double-peaked
\ha\ profile \citep{bb71,dietrich12}, 
a feature that could be a sign of non-virial motions \citep[in particular of accretion disk emission, e.g.][]{eracleous94,eracleous03,
gezari07}.
Therefore our dataset is composed as follows:

\begin{enumerate}
\item H$\beta$ sample. The largest sample considered in this work includes 39 RM AGN with \hb\ FWHM coming either from a mean or a single spectrum.
By requiring that these AGN have an X--ray luminosity measured either in the 2-10 keV or 14-195 keV band 
reduces the sample to 35 objects. 
\item H$\alpha$ sample. Thirty-two\footnote{The AGN having \ha\ FWHM are 33, but we 
excluded the source Mrk 202 as the \ha\ FWHM is deemed to be unreliable \citep{bentz10}.} AGN
have \ha\ FWHM. 
Among this sample, Mrk 877 and SBS 1116+583A do not have an $L_{\rm{X}}$ measurement available. 
Thus the final sample having both \ha\ FWHM and X--ray luminosity includes 30 galaxies.
\item NIR sample. The FWHM of the NIR emission lines were taken from 
\citet[i.e. 19 \pa, 20 \pb\ and 16 \hei]{landt08,landt13}.
We added the measurements of NGC 3783 that we observed simultaneously in 
the ultraviolet, optical and NIR with Xshooter \citep{onori16inpress}.
Therefore the total NIR sample, with available X--ray luminosity, counts: 19 \pa, 21 \pb\ and 17 \hei.
\end{enumerate}
The details of each RM AGN are reported in Table \ref{tab:1}. 
The intrinsic hard X--ray luminosities have been taken 
either from the SWIFT/BAT 70 month catalogue \citep[14-195 keV, $L_{14-195\,\rm{keV}}$;][]{baumgartner13}, 
or from the CAIXA catalogue \citep[2-10 keV, $L_{2-10\,\rm{keV}}$;][]{bianchi09}. 
PG 1411+442 has public 2-10 keV luminosity from \citet{piconcelli05}, 
while
Mrk 1310 and NGC 4748 have public XMM observations, therefore we derived 
their 2-10 keV luminosity via X--ray spectral fitting. 
Both X--ray catalogues list the 90\% confidence level uncertainties on the $L_{\rm{X}}$ and/or on the hard X--ray fluxes, 
which were 
converted into the 1$\sigma$ confidence level. 
For PG 1411+442, as the uncertainty on the $L_{2-10\,\rm{keV}}$ has not been published \citep{piconcelli05}, 
a 6\% error (equivalent to 10\% at the 90\% confidence level) has been assumed.
All the FWHMs listed in Table \ref{tab:1} have been corrected for the instrumental resolution broadening.
When possible, we always preferred to use coeval (i.e. within few months) FWHM measurements of the NIR and optical lines. 
This choice is dictated by the aim of verifying whether the optical and NIR emission lines
are originated at similar distance in the BLR.
The virial masses have been 
taken mainly from the compilations of \citet{grier13} and \citet{hk14}, 
and were computed from the $\sigma_{\rm{line}}$(\hb), measured from the root-mean-square (rms) 
spectra taken during RM campaigns, and the updated \hb\ time lags \citep[see][]{zu11, grier13}.
For 3C 273 and Mrk 335 
the logarithmic mean of the measurements available 
\citep{hk14} were used.

To the data presented in Table \ref{tab:1}, we also 
added six AGN1 from \citet{landt08,landt13} (see Table \ref{tab:1b})
without neither RM $M_{\rm{BH}}$ nor
bulge classification, that however have simultaneous measurements of 
optical and/or NIR lines. 
They are, namely: 
H 1821+643, H 1934-063, H 2106-099, HE 1228+013, IRAS 1750+508, PDS 456. 
Table \ref{tab:1b} also lists the optical data of Mrk 877 and SBS 1116+583A.
These additional eight sources that do not appear in Table \ref{tab:1} are used 
only in the next Section in which emission line relations are investigated.

\begin{figure*}
	\centering
	\includegraphics[width=0.4\hsize]{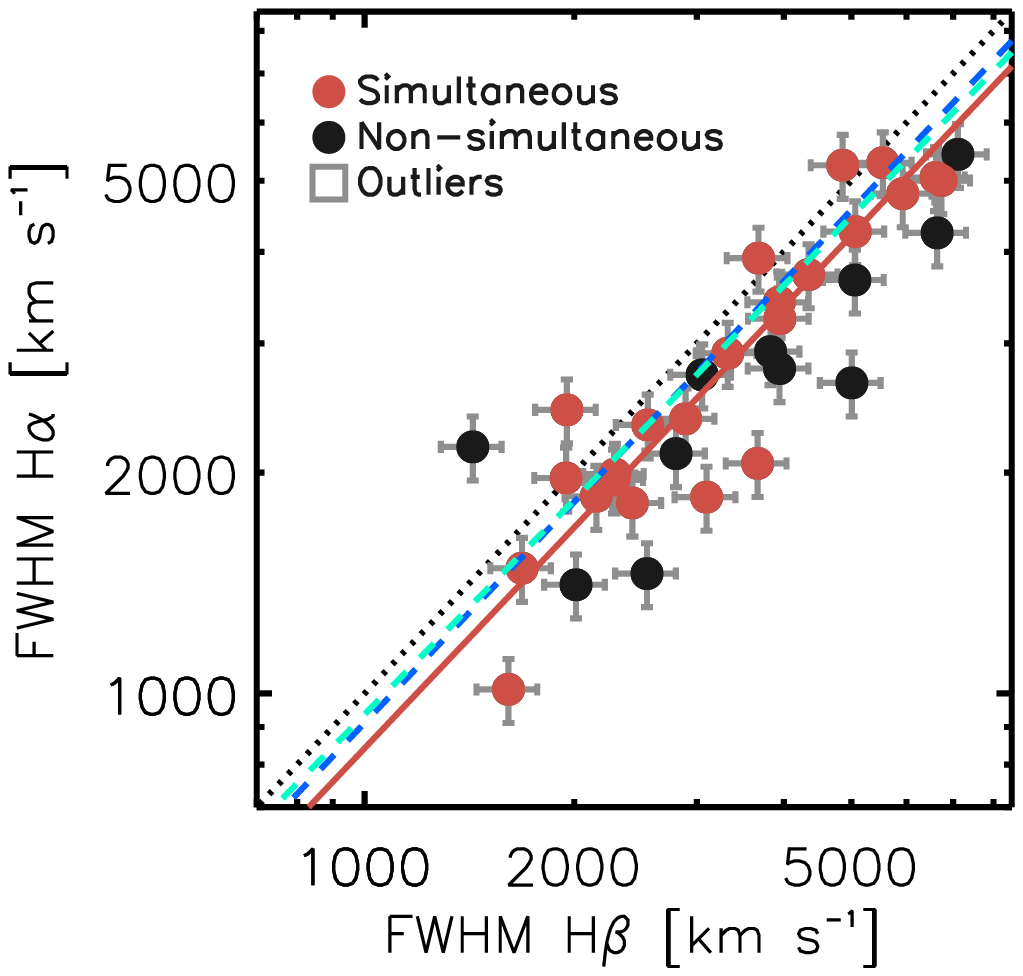}
	\hspace{-1.in}{\includegraphics[width=0.4\hsize]{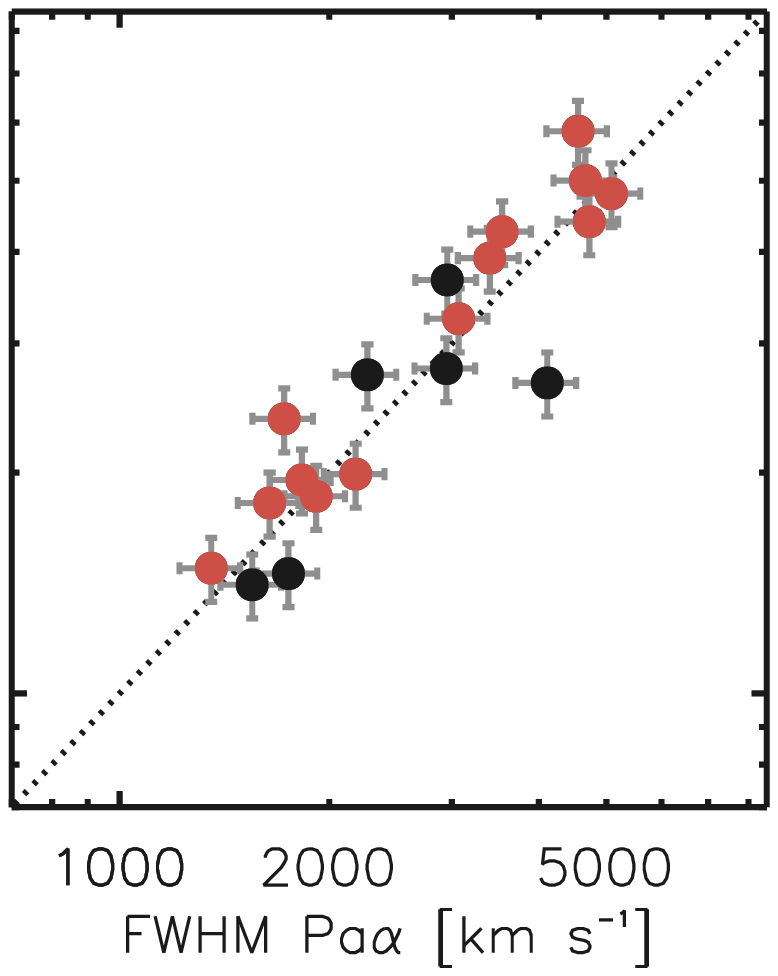}}
	\vspace{-0.25in}{
		\includegraphics[width=0.4\hsize]{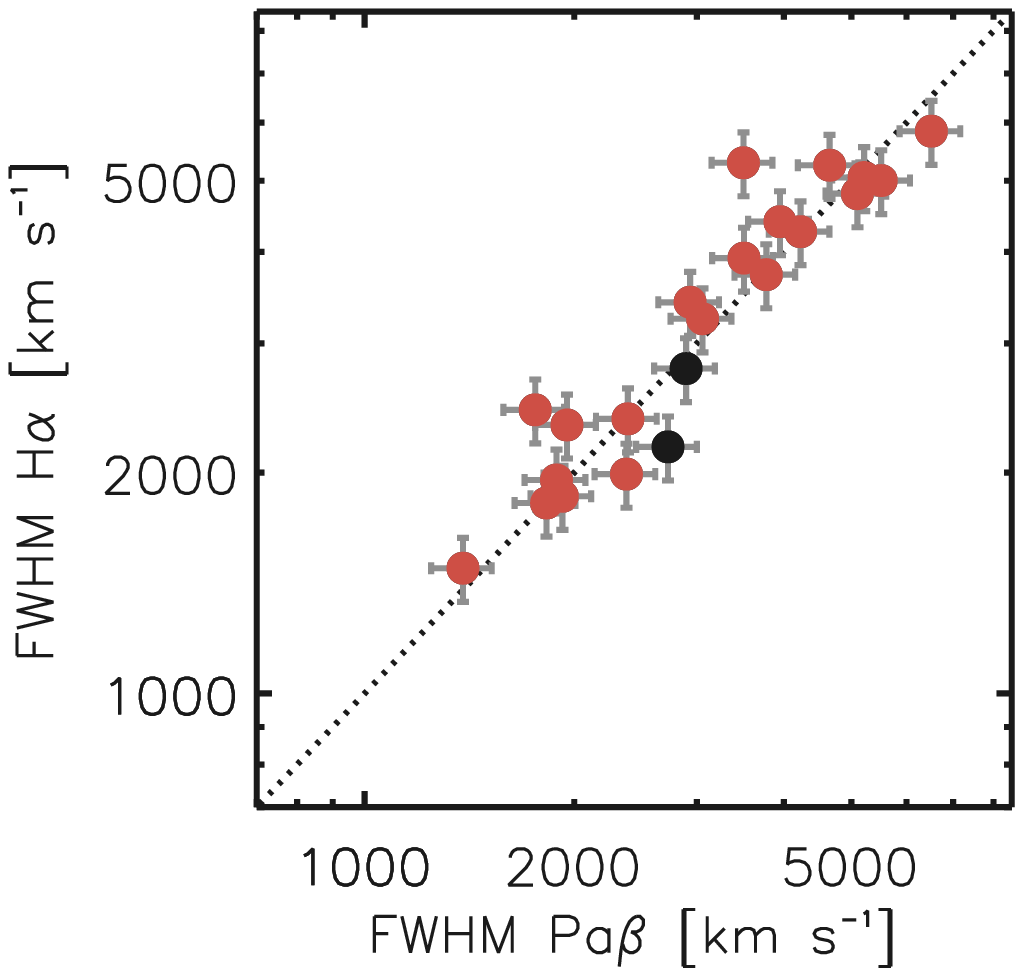}
		\hspace{-1.in}{\includegraphics[width=0.4\hsize]{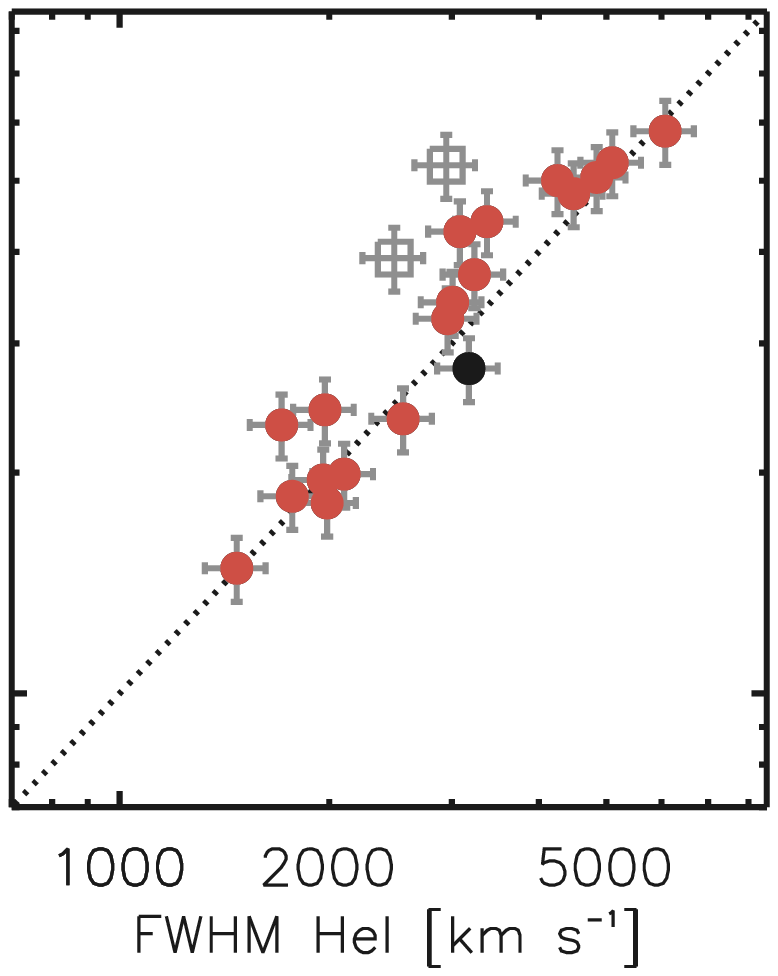}}
	}
	\caption{Linear relations between the FWHM of \ha\ and the FWHM either of \hb, \pa, \pb\ or \hei, 
		from left to right and top to bottom. Red filled circles denote simultaneous observations of the two lines, while 
		black filled circles describe non-simultaneous line measurements. 
		Grey open squares indicate the measurements that are classified as outliers (see text for more details)
		and are not considered in the fits.
		The black dotted line shows the 1:1 relation in all panels. 
		In the top left panel, the 
		best-fit relation computed on the coeval sample is shown as a red solid line. 
		The relations from \citet[dashed cyan]{gh05} and \citet[dashed blue]{mejia16} are also reported. }
	\label{FigFW}%
\end{figure*}

\section{Emission line relations}\label{sec:lines}
 As suggested by several works \citep[e.g.][]{gh05, sl12, mejia16}, 
 the strongest Balmer lines, \ha\ and \hb, seem to come from the same area of the BLR. 
 If we confirm that a linear correlation between 
 \ion{H}{i} and \ion{He}{i}
 optical and NIR lines (i.e. \pa, \pb\ and \hei) does exist, this has two consequences: 
 it will indicate i) that these lines come from 
 the same region of the BLR, and ii) that the widely assumed virialization of \hb\ also implies the virialization of
 \ha\ and of the NIR lines. 

Figure \ref{FigFW} shows the results of our analysis, by comparing 
the FWHM of \ha\ with the FWHMs of the 
\hb , \pa , \pb\ and \hei\ lines. In all panels  
the coeval FWHMs are shown by red filled circles while those not coeval by black filled circles.
Although in some cases the uncertainties are reported in the literature, following the studies of 
\citet{grupe04}, \citet{vp06}, \citet{landt08}, \citet{denney09}, 
we assumed a common uncertainty of 10\% on the FWHM measurements.

The top left panel of Fig. \ref{FigFW} shows the relation between the two Balmer emission lines.
We find a good agreement between the FWHMs
of \ha\ and \hb. 
Using the sub-sample of 23 sources with simultaneous measurements,
the Pearson correlation coefficient results to be $r\simeq 0.92$, with a probability of 
being drawn from an uncorrelated 
parent population as low as $\sim 6\times 10^{-10}$.
The least-squares problem was solved using the symmetrical regression routine 
FITEXY \citep{press07} that can incorporate errors on both variables and allows us to account for intrinsic scatter.
We fitted a log-linear relation to the simultaneous sample and found 
\begin{equation}\label{eq:HaHb}
\log FWHM ({\rm H}\alpha) = \log FWHM ({\rm H}\beta) - (0.075\pm
0.013) \, .
\end{equation} 

The above relation means that \hb\ is on average 0.075 dex broader than \ha, with a scatter of $\sim$0.08 dex.
This relation has a reduced $\chi^2_{\nu}\simeq1.68$. 
We performed the F-test to verify the significance of this non-zero offset with respect to a 1:1 relation, 
getting a probability value of $\sim$2e-4 
that the improvement of the fit was obtained by chance\footnote{Throughout this work in the F-test we use a threshold  
of 0.012, corresponding to a 2.5$\sigma$ Gaussian deviation, in order to rule out 
the introduction of an additional fitting parameter.}.
Therefore in this case the  
relation with a non-zero 
offset resulted to be highly significant.
We also tested whether this offset changes according to the bulge 
classification, when available. No significant difference was found, as the offset of the pseudo bulges resulted to 
be $0.076 \pm 0.020$ and for the classical/elliptical $0.074 \pm 0.020$.

Equation \ref{eq:HaHb} is shown as a red solid line in the top left panel of Fig. \ref{FigFW}.
Our result is in fair agreement (i.e. within 2$\sigma$) with other independent estimates, that are shown as cyan \citep{gh05}
and blue \citep{mejia16} dashed lines. 
If instead we consider the total sample of 34 AGN 
having both \ha\ and \hb\ measured (i.e. including also non-coeval FWHMs), 
we get an average offset of $0.091 \pm 0.010$ 
with a larger scatter ($\sim$0.1 dex). 
Also in this case, the offset does not show a statistically significant dependence 
on the bulge classification, as the offset of elliptical/classical bulges resulted to be $0.102\pm0.014$ 
and for pseudo bulges it resulted to be $0.080\pm0.019$.
In all the aforementioned fits, three outliers\footnote{The outlier values are marked with a dag in Table \ref{tab:1}.} have been excluded even though the 
FWHMs were measured simultaneously. The excluded galaxies namely are
Mrk 1310, Mrk 590 and NGC 5548.
The first one has been excluded because the \ha\ measurement is highly uncertain 
\citep[$561^{+960}_{-136}$ km s$^{-1}$,][]{bentz10},
the latter two have extremely broader \hb\ than either \ha, \pa, \pb\ and \hei. 
This fact is due to the presence of a prominent ``red shelf'' in the \hb\ of these two sources \citep{landt08}.
This red shelf is also most likely responsible for the average trend 
observed between \hb\ and \ha, i.e. of \hb\ being on average broader than \ha\ (Equation \ref{eq:HaHb}).
Indeed it is well known \citep[e.g.][]{derobertis85, marziani96, marziani13} that 
the \hb\ broad component 
is in part blended with weak 
\ion{Fe}{ii} multiplets, \heii\ $\lambda$4686 and 
\hei\ $\lambda$4922, 5016 \citep{veron02,kollatschny01}.
The simultaneous sample gives a relation with lower scatter than 
the total sample. Indeed, the non-simultaneous measurements introduce 
additional noise due to the well-known AGN variability phenomenon. Therefore 
in the following Sections we will use the average offset between the FWHM of \ha\ and \hb\ computed using the coeval sample
(i.e. Equation \ref{eq:HaHb}), 
which also better agrees with the relations already published by \citet{gh05} and \citet{mejia16}.

\begin{table*}
	\caption{Results of the fits of the virial relations.}             
	\label{tab:2}
	\centering                          
	\begin{tabular}{l c c c c c c c}
				
		\hline\hline 
		\noalign{\smallskip}
		\multicolumn{8}{c}{$M_{\rm{BH}}$ vs $VP(\langle FWHM({\rm H}\alpha) \rangle , \langle L_{2-10 {\rm \, keV}}\rangle )$}\\
		\noalign{\smallskip}
		\hline
		\noalign{\smallskip}
		sample & $a$&$b$ & N & r & Prob(r) & $\epsilon_{obs}$ & $\epsilon_{intr}$ \\
		(1) & (2) & (3) & (4) & (5) & (6) & (7)&(8) \\
		\noalign{\smallskip}
		\hline
		\noalign{\smallskip}
		All      &8.032 $\pm$    0.014& 1\tablefoottext{a} & 37 & 0.838 & 9$\times$10$^{-11}$ &  0.40 & 0.38 \\
		Clas     &8.083 $\pm$    0.016& 1\tablefoottext{a} & 23 & 0.837 & 7$\times$10$^{-7}$  &  0.38 & 0.37 \\
		Pseudo 	 &7.911 $\pm$    0.026& 1\tablefoottext{a} & 14 & 0.731 & 3$\times$10$^{-3}$  &  0.40 & 0.38 \\
		All\tablefoottext{b}    &8.187 $\pm$  0.021&  1.376 $\pm$ 0.033 & 37 & 0.831 & 2$\times$10$^{-10}$ &   0.49 & 0.48 \\
		\noalign{\smallskip}
		\hline
		
	\end{tabular}
	
	\tablefoot{ Best fitting parameters of the virial relations (see Equation \ref{eq:cal}) between the 
		$M_{\rm{BH}} = f \times M_{\rm{vir}}$, with $\langle f\rangle=4.31$ \citep{grier13}, and
		the average VP given by the mean FWHM
		(once the \hb\ has been converted 
		into \ha) and the mean $L_{2-10 \, {\rm keV}}$ (using Equation \ref{eq:Lx} to convert $L_{14-195\,{\rm keV}}$).
		Columns are: (1) sample bulge type, 
		(2) and (3) zero point and slope of the virial relation,
		 (4) number of objects of 
		each sample, (5) and (6) Pearson correlation coefficient with its t-student probability,
		(7) logarithmic spread of the data on the $M_{\rm{BH}}$ axis, 
		(8) intrinsic logarithmic spread of the data on the y axis as before.\\
		\tablefoottext{a}{Fixed value.} \\
		\tablefoottext{b}{In this sample different virial factors for 
				classical/elliptical and pseudo bulges have been used: $f_{\rm{CB}}=6.3$, $f_{\rm{PB}}=3.2$ \citep{hk14}.} 
		}
	
\end{table*}

The other three panels of Fig. \ref{FigFW} show the relations between the \ha\ and the NIR 
emission lines \pa\ (19 objects) , \pb\ (22) and \hei\ (19).
When compared to \ha, the samples have 
Pearson correlation coefficient $r$ of 0.92, 0.94 and 0.95, with probabilities of 
being drawn from an uncorrelated 
parent population as low as $\sim 2\times 10^{-8},  9 \times 10^{-11}$ and  $ 5\times 10^{-10}$
for the \pa, \pb\ and \hei, respectively. 
No significant difference is seen between the emission line widths of \ha\ and the NIR lines. 
This is not surprising as already \citet{landt08} noted that there was a good agreement between the 
FWHM of the \pb\ and the two strongest Balmer lines, 
though an average trend of \hb\ being larger than \pb\ was suggested (a quantitative analysis was not carried out).
We fitted log-linear relations to the data and always found that the 1:1 relation is the best 
representation of the sample. 
We found a reduced $\chi^2_\nu$ of 1.54, 1.12 and 1.14 for 
\pa, \pb\ and \hei, respectively.
The F-test was carried out in order to quantitatively verify whether 
the equality relation is preferred with respect to relations either with a non-zero offset or 
also including a free slope. 
The improvements with a relation having free slope
resulted not to be highly significant, and therefore the more physically motivated 1:1 relation was preferred.
These best-fitting relations are shown as black dotted lines in the remaining three panels of Figure \ref{FigFW}.
The relation between \ha\ and the \pb\ emission line has been fitted using the whole sample,
while for the \pa\ and \hei\ correlations we excluded the sources: NGC 7469 (for the \pa), Mrk 79 and NGC 4151 
(for the \hei ; shown as grey open squares in the bottom right panel of Fig. \ref{FigFW}).
Although these three sources have simultaneous optical and NIR observations \citep{landt08}, all 
show significantly narrower width of the \pa\ or \hei\ than all the other available 
optical and NIR emission lines.

\begin{figure*}
	\centering
	\includegraphics[width=0.4\hsize]{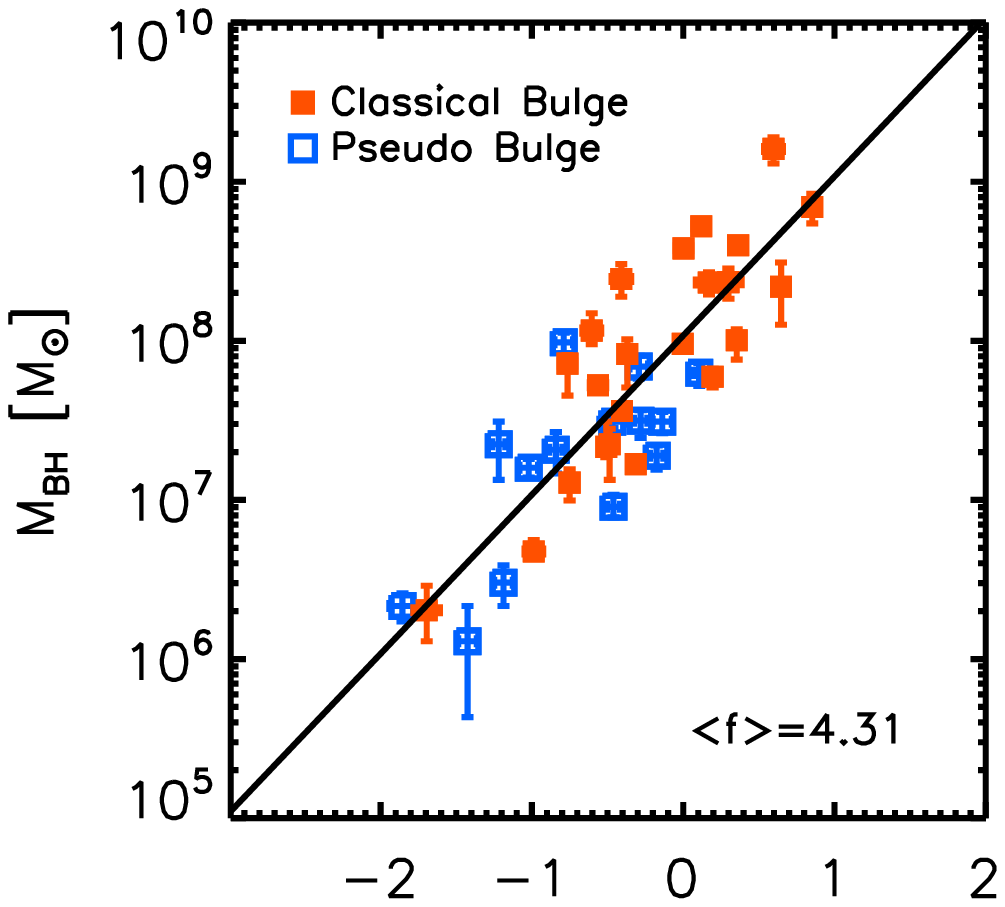}
	\hspace{-1.in}{\includegraphics[width=0.4\hsize]{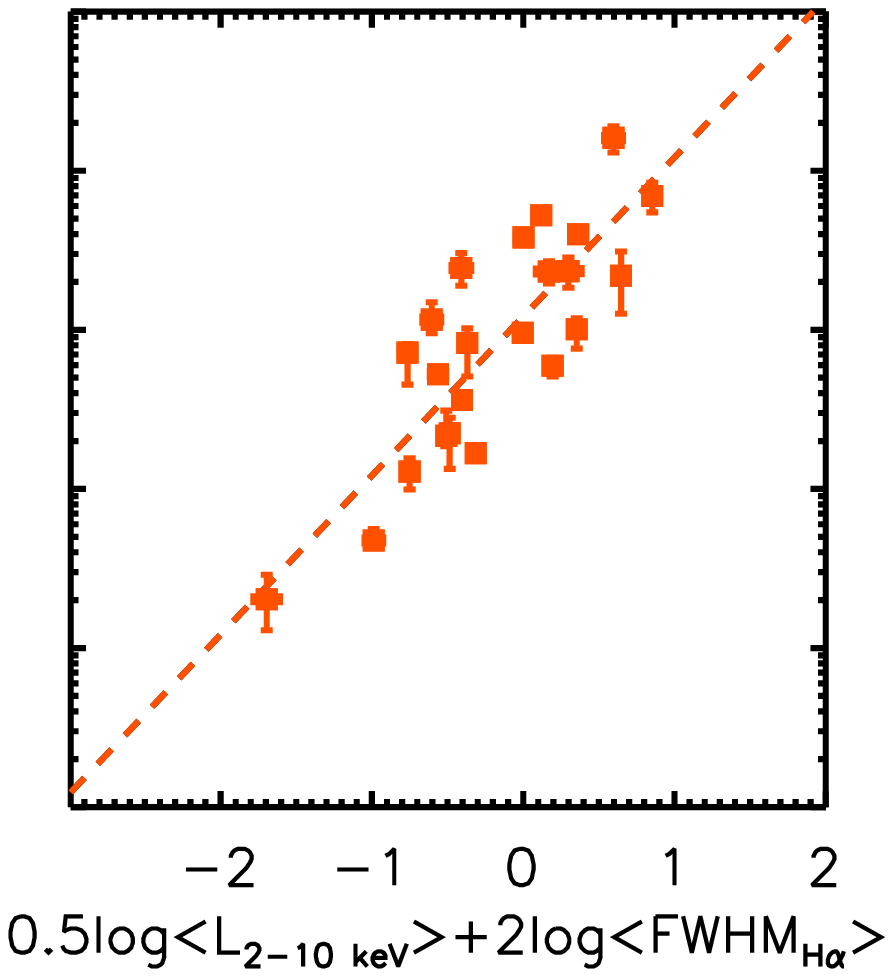}}
	\hspace{-1.in}{\includegraphics[width=0.4\hsize]{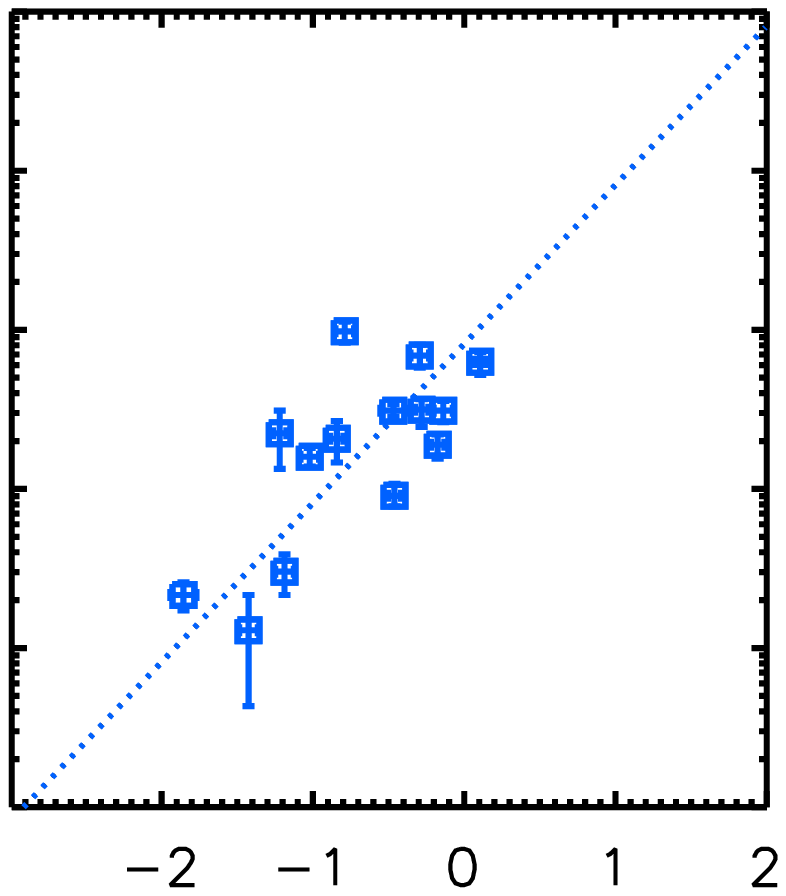}}\\
	
	\vspace{-0.25in}{	
		\includegraphics[width=0.4\hsize]{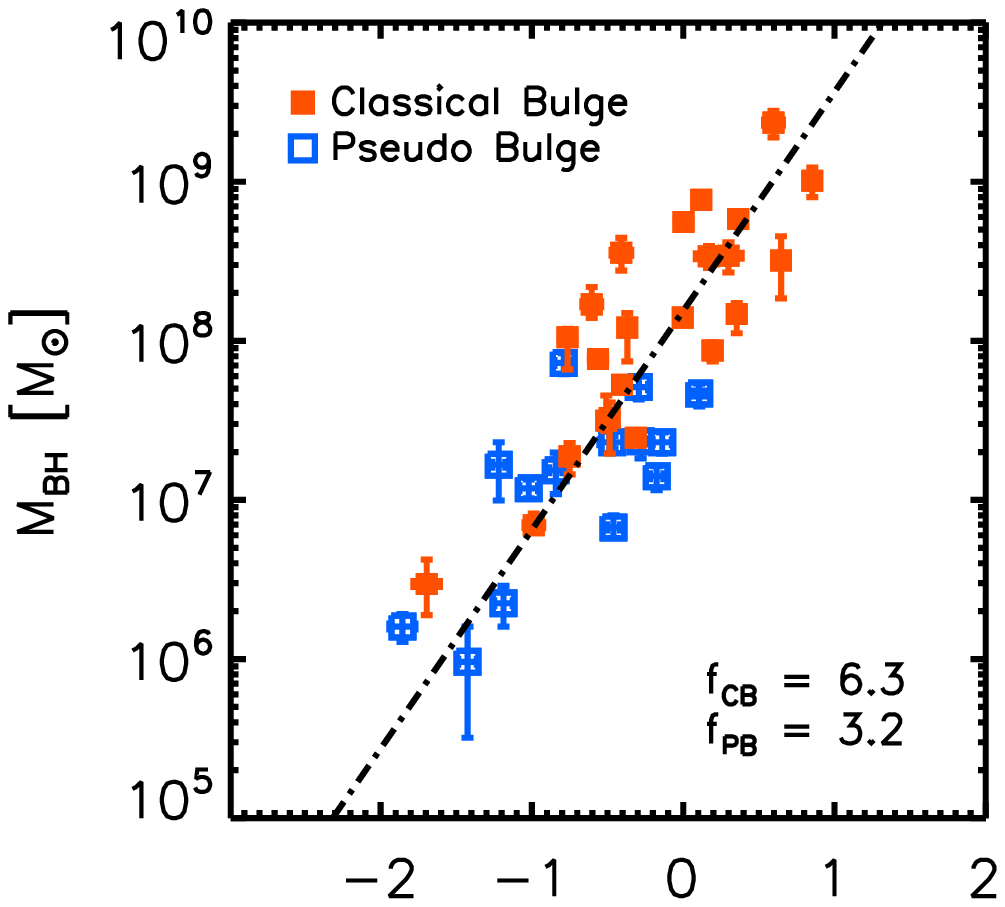}
		\hspace{-1.in}{\includegraphics[width=0.4\hsize]{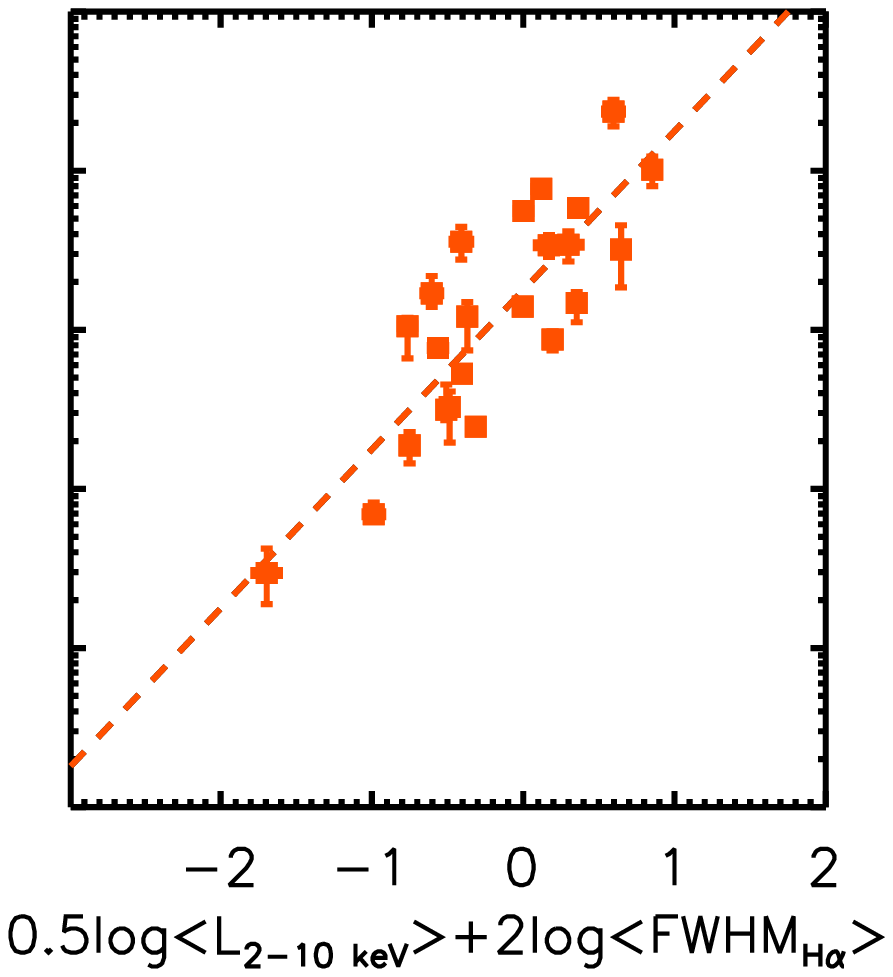}}
		\hspace{-1.in}{\includegraphics[width=0.4\hsize]{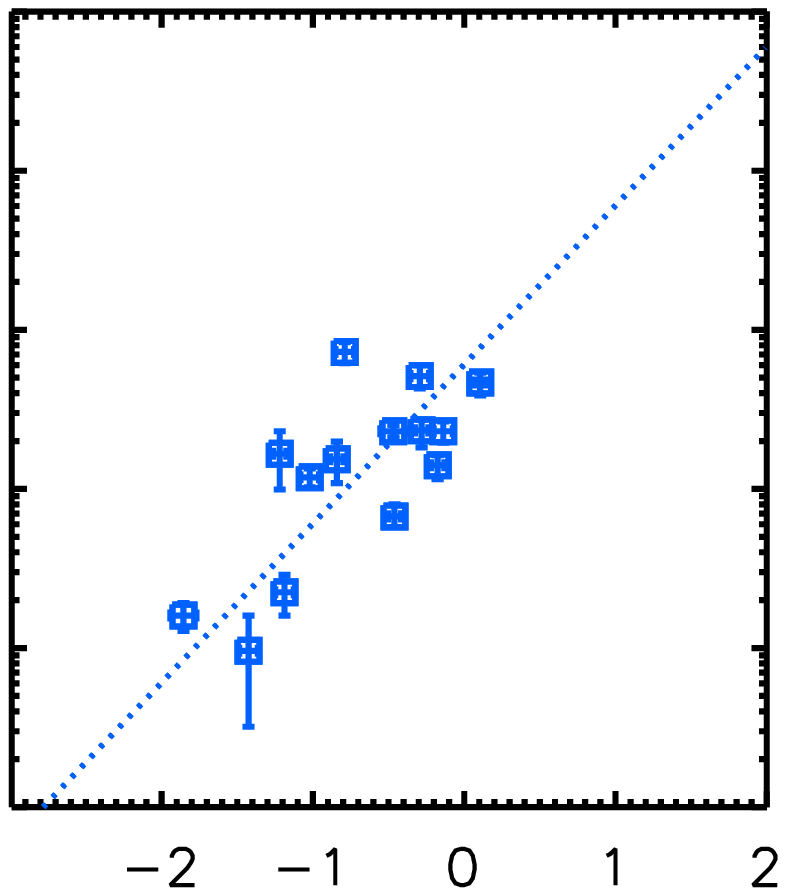}} }
	\caption{Virial relations between the BH mass $M_{\rm{BH}} = f \times M_{\rm{vir}}$ and 
		the average VP given by the mean FWHM (once the \hb\ has been converted 
		into \ha) and the mean $L_{2-10 \, {\rm keV}}$ (using Equation \ref{eq:Lx} to convert $L_{14-195\,{\rm keV}}$).
		In the top panels the black hole masses have been calculated assuming 
		$\langle f \rangle = 4.31$ \citep{grier13}, while in the bottom panels 
		two different $f$-factors $f_{\rm{CB}} = 6.3$ for classical bulges and 
		$f_{\rm{PB}} = 3.2$ for pseudobulges \citep{hk14}
		have been adopted to determine $M_{\rm{BH}}$.
		All the VPs are normalized as specified in Equation \ref{eq:cal} (see Tables \ref{tab:2}-\ref{tab:3} 
		for the resulting best-fit parameters). 
		In the left panels the total calibrating sample is shown, 
		while in the middle and right panels the sub-samples of classical (red filled squares)
		and pseudobulges (blue open squares) are shown separately.
		The lines show the best-fitting virial relations derived for each sample.
	}
	\label{FigVP} 
\end{figure*}

\section{Virial mass calibrations}\label{sec:cal}
In the previous Section we showed that the \ha\ FWHM is 
equivalent to the widths of the NIR emission lines, \pa, \pb\ and \hei,
while it is on average 0.075 dex narrower than \hb.  
In order to minimize the uncertainties on the estimate of the 
zero point and slope that appear in Equation \ref{eq:cal}
we used the whole dataset listed in Table \ref{tab:1}.
This is possible because, besides the linear 
correlations between the optical and NIR FWHMs,
also the intrinsic hard X--ray luminosities $L_{2-10 \, \rm{keV}}$ and $L_{14-195\,\rm{keV}}$ are correlated. 
Indeed, as expected in AGN, we found in our sample a relation between the two hard X--ray luminosities,
\begin{equation}\label{eq:Lx}
\log L_{2-10 \, \rm{keV}} = \log L_{14 - 195 \, \rm{keV}} - (0.567 \pm 0.004) \, ,
\end{equation}
which corresponds to an average X--ray photon index $\langle \Gamma \rangle \simeq 1.67$ ($f_\nu \propto \nu ^{-(\Gamma-1)}$).
We can therefore calculate, for each object of our sample, a sort of average VP, 
which has been computed by using the average FWHM
of the emission lines (the \hb\ has been converted into \ha\ by using Equation \ref{eq:HaHb}) and 
the average X--ray luminosity (converted into 2-10 keV band using Equation \ref{eq:Lx}).
When computing the average FWHM, the values that in the 
previous Section were considered outliers were again excluded.
However we note that each RM AGN has at least one valid FWHM measurement, 
therefore none of the AGN has been excluded. 
This final RM AGN sample is the largest with available bulge classification \citep{hk15} and 
hard $L_{\rm{X}}$ and counts a total of 37 sources, 23 of which are elliptical/classical and 14 
are pseudo bulges\footnote{This sample is made of all sources having hard X--ray luminosity 
measurements among those of \citet{hk14}, whose sample of bulge-classified RM AGN
includes $\sim$90\% of all the RM black hole masses available in the literature.}. 

We want to calibrate the linear virial relation given in Equation \ref{eq:cal},
where $M_{\rm{BH}}$ is the RM black hole mass, which is equal to $f \times M_{\rm{vir}}$,
$a$ is the zero point and $b$ is the slope of the average VP.
We fitted Equation \ref{eq:cal} for the whole sample of RM AGN assuming 
one of the most updated virial factor $\langle f \rangle = 4.31$ \citep{grier13}, which 
does not depend on the bulge morphology. The data 
have a correlation coefficient $r=0.838$ which corresponds to a probability as low as $\sim9\times10^{-11}$ 
that they are randomly extracted from an uncorrelated parent population.
As previously done in Sect. \ref{sec:lines}, we performed a 
symmetrical regression fit using FITEXY \citep{press07}.
We first fixed the slope $b$ to unity   
finding the zero point $a=8.032\pm 0.014$. 
The resulting observed spread $\epsilon_{\rm{obs}}$ is 0.40 dex, while the 
intrinsic spread $\epsilon_{\rm{intr}}$ (i.e. once the contribution from the data uncertainties 
has been subtracted in quadrature) results to be 0.38 dex.
We also performed a linear regression, allowing the slope $b$ to vary. 
The F-test was carried out to quantitatively verify whether our initial assumption 
of fixed slope had to be preferred to a relation having free slope.
The F-test gave a probability of $\sim0.07$ which is 
not significantly small enough to demonstrate that 
the improvement using a free slope is not obtained by chance. 
Therefore, the use of the more physically motivated
relation (having slope $b=1$) that depends only linearly on the VP was preferred. The resulting best-fitting parameters are reported in Table \ref{tab:2}, while the virial 
relation is shown in the top-left panel of Fig. \ref{FigVP} (black solid line).\\

We then splitted the sample into elliptical/classical (23) and pseudo (14) bulges, adopting 
the same virial factor $\langle f \rangle=4.31$.
The two samples have correlation coefficient $r>0.7$ 
with probabilities lower than $\sim 10^{-3}$ that 
the data have been extracted randomly from an 
uncorrelated parent population (see Table \ref{tab:2}). 
Again we first fixed the slope $b$ to unity, obtaining
the zero points $a=8.083\pm0.016$ and $a=7.911\pm0.026$
for classical and pseudo bulges, respectively.  
We also performed a linear regression allowing a free slope. 
The F-test was carried out 
and gave probabilities greater than 0.05 for both the classical and 
the pseudo bulges samples. Therefore the more physically motivated
relations that depend linearly on the VP were preferred, as previously found for 
the whole sample. Top middle and top right panels of Fig. \ref{FigVP} show the resulting best-fit 
virial relations for classical (in blue) and pseudo bulges (in red).
It should be noted that the average difference between the zero points $a$ of the two bulge type populations 
is $\sim$0.2 dex.\\
Obviously, the same fitting results, using these two sub-samples separately,
are obtained if the 
recently determined different $f$ factors of 6.3 and 3.2 for classical 
and pseudo bulges \citep{hk14} are adopted. 
However, as expected, the difference between the zero points of the 
two populations becomes larger ($\sim$0.5 dex)
as the zero points result to be $a=8.248\pm0.016$ and $a=7.782\pm0.026$
for the classical and pseudo bulges, respectively.\\
Finally we performed a calibration of Equation \ref{eq:cal} 
for the whole sample adopting the two virial factors $f_{\rm{CB}}=6.3$ and $f_{\rm{PB}}=3.2$,
according to the bulge morphological classification. The data result to have a correlation coefficient $r=0.831$ with a 
probability as low as $\sim 10^{-10}$ to have been drawn 
randomly from an uncorrelated parent population.
As previously described, we proceeded fixing the slope $b$ to unity and then 
fitting a free slope. The F-test  
gave a probability lower than 0.01, therefore the solution with 
slope\footnote{It should be noted that 
a slope different than unity implies that the $L_X-R$ relation (R$\propto$L$^\alpha$)
has a power $\alpha \ne 0.5$, contrary to what found by \citet{greene10}.} $b=1.376\pm0.033$ was in this case considered statistically significant (see Table \ref{tab:2}).
	The observed and intrinsic spreads resulted to be $\sim$0.5 dex (see Table \ref{tab:2}).
	The bottom panels of Fig. \ref{FigVP} show the virial relations that we obtained 
	for the whole sample (left panel, black dot-dashed line) and separately for classical (middle panel,
	red dashed line) and pseudo bulges (right panel, blue dotted line), once the 
	two different virial factors $f$ are adopted according to the bulge morphology.

We note that 
it is possible to convert our 
virial calibrations (which were 
estimated using the mean line widths, once converted into the \ha\ FWHM, and 
the mean X--ray luminosity, once converted into the $L_{2-10 \, \rm{keV}}$) into 
other equivalent relations based on either the \hb, \pa, \pb\ and \hei\ FWHM 
and the $L_{14-195 \, \rm{keV}}$, by using the correlations shown in Equations \ref{eq:HaHb} and 
\ref{eq:Lx}. To facilitate the use of our virial BH mass estimators, we list in Table \ref{tab:3} how the 
virial zero point $a$ changes according to the couple of variables that one wish to use. 
Moreover, as shown in the virial relation in the top part of Table \ref{tab:3},
it is possible to convert the resulting BH masses for different assumed virial $f$ factors 
adding the term $\log(f/f_0)$, where $f_0$ is the virial factor that was assumed when each sample was fitted. 
The values of $f_0$ are also reported in Table \ref{tab:3} for clarity.
Note that, in the last case where a solution was found for the 
total sample using separate $f$-factors for classical and 
pseudobulges, the $\log (f/f_0)$ correction cannot be used. 
However, this last virial relation 
is useful in those cases where
the bulge morphological type is unknown
and one wish to use a solution which 
takes into account the two virial factors 
for classical and pseudobulges as measured by \citet{hk14}. 
Otherwise, the virial BH mass estimator calculated 
fitting the whole dataset assuming a 
single $\langle f \rangle = 4.31$ can be used (and if necessary converted adopting a
different average $f$ factor).

\begin{table*}
	\caption{Final virial BH mass estimators.	}             
	\label{tab:3}
	\centering                          
	\begin{tabular}{r c c c c c c c c c c c}
		\hline\hline 
		\noalign{\smallskip}
		\multicolumn{12}{c}{$\log \left(\dfrac{M_{\rm{BH}}}{M_\odot}\right)  = a + b \left[ 2 \log \left(\dfrac{FWHM}{10^4 \, {\rm km \, s^{-1}}}\right) + 0.5 \log \left(\dfrac{L_X}{10^{42} {\rm erg \, s^{-1}}}\right) \right] + \log \left( \dfrac{f}{f_0} \right)$}\\
		\noalign{\smallskip}
		\hline
		\noalign{\smallskip}
		 & \multicolumn{2}{c}{Variables} & & \multicolumn{2}{c}{All ($f_0=4.31$)} & & \multicolumn{2}{c}{CB ($f_0=4.31$)} & & \multicolumn{2}{c}{PB ($f_0=4.31$)} \\
		\noalign{\smallskip}
		\cline{2-3} \cline{5-6} \cline{8-9} \cline{11-12} 
		\noalign{\smallskip}
		& $L_X$& FWHM & & a & b\tablefoottext{a} & & a & b\tablefoottext{a} & & a & b\tablefoottext{a}  \\
		& (1) & (2) & & (3) & (4)\phantom{\tablefoottext{a}} & & (5) & (6)\phantom{\tablefoottext{a}} & & (7) & (8)\phantom{\tablefoottext{a}} \\
		\noalign{\smallskip}
		\cline{2-3} \cline{5-6} \cline{8-9} \cline{11-12}
		\noalign{\smallskip}
	a1 )	& $L_{2-10 {\rm \, keV}}$&     \ha\ (or \pa, \pb, \hei)&  &  8.03 $\pm$ 0.01 & 1\phantom{\tablefoottext{a}} & & 8.08 $\pm$ 0.02 & 1\phantom{\tablefoottext{a}} & & 7.91 $\pm$ 0.03 & 1\phantom{\tablefoottext{a}} \\
	a2 )	& $L_{2-10 {\rm \, keV}}$ &    \hb\				 	 &  &  7.88 $\pm$ 0.03 & 1\phantom{\tablefoottext{a}} & & 7.93 $\pm$ 0.03 & 1\phantom{\tablefoottext{a}} & & 7.76 $\pm$ 0.04 & 1\phantom{\tablefoottext{a}}\\
	a3 )	& $L_{14-195 {\rm \, keV}}$& 	 \ha\ (or \pa, \pb, \hei)&  &  7.75 $\pm$ 0.01 & 1\phantom{\tablefoottext{a}} & & 7.79 $\pm$ 0.02 & 1\phantom{\tablefoottext{a}} & & 7.63 $\pm$ 0.03 & 1\phantom{\tablefoottext{a}}\\
	a4 )	& $L_{14-195 {\rm \, keV}}$ &	 \hb\				 	 &  &  7.60 $\pm$ 0.03 & 1\phantom{\tablefoottext{a}} & & 7.65 $\pm$ 0.03 & 1\phantom{\tablefoottext{a}} & & 7.48 $\pm$ 0.04 & 1\phantom{\tablefoottext{a}}\\
		& &  & &  &  & &  &  & &  &   \\
		\noalign{\smallskip}	
		\hline	
		\noalign{\smallskip}
		 & \multicolumn{2}{c}{Variables} & & \multicolumn{2}{c}{All\tablefoottext{b}} & & \multicolumn{2}{c}{CB ($f_0=6.3$)} & & \multicolumn{2}{c}{PB ($f_0=3.2$)} \\
		\noalign{\smallskip}		
		\cline{2-3} \cline{5-6} \cline{8-9} \cline{11-12} 
		\noalign{\smallskip}
	 &	$L_X$& FWHM & & a & b & & a & b\tablefoottext{a} & & a &b\tablefoottext{a}\\

b1 ) &	$L_{2-10 {\rm \, keV}}$&     \ha\ (or \pa, \pb, \hei)&  & 8.19 $\pm$ 0.02 & 1.38 $\pm$ 0.03 & & 8.25 $\pm$ 0.02 & 1\phantom{\tablefoottext{a}} & & 7.78 $\pm$ 0.02 & 1\phantom{\tablefoottext{a}} \\
b2 ) &	$L_{2-10 {\rm \, keV}}$ &    \hb\				 	 &  & 7.98 $\pm$ 0.04 & 1.38 $\pm$ 0.03 & & 8.10 $\pm$ 0.03 & 1\phantom{\tablefoottext{a}} & & 7.63 $\pm$ 0.04  & 1\phantom{\tablefoottext{a}}\\
b3 ) &	$L_{14-195 {\rm \, keV}}$& 	 \ha\ (or \pa, \pb, \hei)&  & 7.80 $\pm$ 0.02 & 1.38 $\pm$ 0.03 & & 7.96 $\pm$ 0.02 & 1\phantom{\tablefoottext{a}} & & 7.50 $\pm$ 0.03  & 1\phantom{\tablefoottext{a}}\\
b4 ) &	$L_{14-195 {\rm \, keV}}$ &	 \hb\				 	 &  & 7.59 $\pm$ 0.04 & 1.38 $\pm$ 0.03 & & 7.81 $\pm$ 0.03 & 1\phantom{\tablefoottext{a}} & & 7.35 $\pm$ 0.04  & 1\phantom{\tablefoottext{a}}\\

\hline				
	\end{tabular}
	\tablefoot{Parameters to be used in the virial relation described in the top part of this table. 
			The resulting BH mass values can be converted 
			assuming other virial factor $f$ using the 
			additional term $\log (f/f_0)$. The assumed $f_0$ in each sample is also reported. 
		All the above virial calibrations have an intrinsic spread of $\sim$0.5 dex that should be 
		taken into account when evaluating the accuracy of the BH mass estimates (see Table \ref{tab:2}).
		Columns are: (1) the hard X--ray luminosity, (2) the FWHM, both variables needed to compute the VP, 
		(3) to (8) the zero points $a$ and the slopes $b$ of each sample.\\
		\tablefoottext{a}{Fixed value.}\\
		\tablefoottext{b}{Note that in this sample different $f$ factors, 
			according to the bulge morphology, have been adopted. 
			Therefore the average correction $\log (f/f_0)$ cannot be applied.}
		}	
\end{table*}	

We compared the virial relation derived by \citet{LF15} 
using the \pb\ and the $L_{14-195\, \rm{keV}}$ 
with our two new virial relations, 
which depends on the VP given by the \ha\ FWHM and the $L_{14-195\, \rm{keV}}$,	
obtained using the total sample, and assuming either 1) an average virial factor $\langle f \rangle =4.31$ 
\citep[as used in][]{LF15}
or 2) the two different $f$, 
separately for classical and pseudobulges.	
It results that our two new virial relations give BH masses 
similar to the relation of \citet{LF15} at $M_{\rm{BH}}\sim10^{7.5}$ M$_\odot$, 
while they predict 0.3 (0.8) dex higher BH masses at $M_{\rm{BH}}\sim10^{8.5}$ M$_\odot$
and 0.2 (0.7) dex 
lower masses at $M_{\rm{BH}}\sim10^{6.5}$ M$_\odot$, 
assuming the average $\langle f \rangle = 4.31$ 
(the two $f$-factors $f_{\rm{CB}} = 6.3$, $f_{\rm{PB}}=3.2$).
These differences are due to
the samples used: our dataset 
includes 15 AGN with
$M_{\rm{BH}}\gtrsim10^8$ M$_\odot$, while in the \citet{LF15} sample 
there are only three, and
at $M_{\rm{BH}}\lesssim10^7$ M$_\odot$ our dataset is a factor two larger.

The same comparison was carried out
using the VP given by the \ha\ FWHM and the $L_{2-10\,{\rm keV}}$
with the analogous relation in \citet{bongiorno14}.
All the relations predict similar masses in the $M_{{\rm BH}} \sim 10^{7.5}$ M$_\odot$ range, 
while our new calibrations give 0.1 (0.2) dex smaller (higher) masses at $M_{{\rm BH}} \sim 10^{8.5}$ M$_\odot$
and 0.2 (0.1) dex bigger (lower) BH masses at $M_{{\rm BH}} \sim 10^{6.5}$ M$_\odot$, assuming the average $\langle f \rangle = 4.31$ 
(the two $f$-factors $f_{\rm{CB}} = 6.3$, $f_{\rm{PB}}=3.2$).

Finally our analysis shows some similarities with the results of \citet{hk15}, 
	who recently calibrated SE optical virial relation based on the \hb\ FWHM and 
	$L_{1500}$, using the total calibrating sample of RM AGN and 
	separated according to the bulge morphology into classical and pseudo bulges.
	They found that in all cases the $M_{\rm{BH}}$ depends on the optical 
	VP with slope $b=1$ and with different zero points $a$ for classical and pseudobulges. 
	This difference implies that BH hosted in pseudo bulges are predicted to be 
	0.41 dex less massive than in classical bulges.
	When we adopt the same $f$-factors used by \citet{hk15}, 
	we do similarly find that the zero point $a$ of classical bulges 
	is $\sim$0.5 dex greater than for pseudo bulges.
	However we do not confirm their result obtained using
	the total sample, as we find that the best-fitting parameter $b$ of our VP 
	should be different than one.
	At variance when the same average $\langle f \rangle =4.31$ is adopted, 
	both in the total and in the sub-samples of classical and 
	pseudo bulges, we find slope $b=1$ relations, while the
 	zero points of classical and pseudo bulges still show an offset of $\sim$0.2 dex.

\section{Discussion and Conclusions}\label{sec:concl}
This work was prompted by the results of \citet{hk15} who have calibrated 
optical different virial relations according to the bulge morphological classification into 
classical/elliptical and pseudo bulges \citep{hk14}.
In order to provide virial relations to be used 
also for moderately absorbed AGN, 
following \citet{LF15},
we extended the approach of 
\citet{hk15} using the intrinsic hard X--ray luminosity and 
NIR emission lines. 
We thus obtained similar virial relations for the two bulge 
classes but with an offset between the two zero points of $\sim$0.2 dex 
if the same average $\langle f\rangle=4.31$ is used.
If instead two different virial factors $f_{\rm{CB}}=6.3$ and $f_{\rm{PB}}=3.2$
are assumed, the offset becomes linearly larger by a factor of $\sim$2,
confirming the results by \citet{hk15}.
Neglecting the morphological information 
leads to a systematic uncertainty of $\sim$0.2-0.5 dex, that is the 
difference we observe when we split the sample according to the host bulge type.
This uncertainty will be difficult to eliminate  
because of the current challenges at play when attempting to 
accurately measure the properties of the host, 
especially at high redshift and/or for luminous AGN.
As already stated by \citet{hk15}, AGN with  
$M_{\rm{BH}} \gtrsim 10^8$ M$_\odot$ are most probably 
hosted by elliptical or classical bulges, as suggested 
also by the current BH mass measures in inactive galaxies \citep[e.g.][]{hk14}.
Similarly, $M_{\rm{BH}} \lesssim 10^6$ M$_\odot$ are very likely hosted 
in pseudo bulges \citep[e.g.][]{greene08, jiang11}.
However, the two populations significantly overlap in the range $10^6\lesssim M_{\rm{BH}} / \rm{M}_\odot \lesssim 10^8$
and therefore without bulge classification the BH mass estimate is accurate only within a factor of $\sim$0.2-0.5 dex.
Probably accurate bulge/disk decomposition will be available also 
for currently challenging sources once extremely large telescopes such the EELT become 
operative for the community. 
Indeed the high spatial resolution that can be achieved with sophisticated 
multiple adaptive optics will enable to probe scales of 
few hundreds of parsecs in the centre of galaxies at $z\sim2$ \citep{gullieuszik16}.

Obviously, the above results depend on the 
bulge morphological classification. As discussed in the 
introduction, this classification should be carried out carefully, and 
the reliability increases 
by enlarging the 
number of selection criteria used 
\citep{kormendyho13,kormendy16}. It should also be noted that according to some authors 
the main selection criterion should instead be based on the presence or 
not of a bar 
\citep{graham09, graham14, savorgnan15}. 
This simpler selection criterion,
which avoids the difficulty arising from the observation that some 
(at least 10\%) galaxies host both a pseudobulge and a classical bulge 
\citep{erwin03, erwin15},	
is supported by 
dynamical modelling studies by 
\citet{debattista13} and \citet{hartmann14}. 
As a matter of fact it is interesting to remark that 
an offset of 0.3 dex is also observed in the $M_{\rm{BH}} -\sigma_\star$ diagram 
when the galaxies are divided into barred and unbarred \citep[e.g.][]{graham08, graham11, grahamscott13}.
Moreover, \citet{hk14} note that although the presence of a bar does not 
correlate perfectly with bulge type, the systematic difference in 
$f$ between barred and unbarred galaxies qualitatively resembles the dependence 
on bulge type that they found.

Recently, \citet{shankar16} claimed that all
	the previously computed $f$ factors could have been artificially 
	increased by a factor of at least $\sim$3 because of a presence 
	of a selection bias in the calibrating samples, in favour of the more massive 
	BHs.
	This result would imply that all the previous estimate of the virial 
	relations, including those presented in this work, suffer 
	from an almost average artificial offset. 
	If, as discussed by \citet{shankar16}, the offset is not 
	significantly dependent on $M_{\rm{BH}}$, 
	then it is sufficient to rescale our results by a correction factor 
	$\log(f/f_0)$. The same correction term can also be 
	used to convert our relations assuming virial factors
	different than those used in this work.

By testing whether the \ha\ probes a velocity field in the BLR consistent 
with the \hb\ and the other NIR lines \pa, \pb\ and \hei, 
we widened the applicability of our proposed virial relations.
Indeed assuming the virialization of the clouds emitting the \hb\ 
implies the virialization also of the other lines considered in this work. 
Moreover, these lines can be valuable tools to estimate the velocity of 
the gas residing in the BLR also for intermediate (e.g. Seyfert 1.9) and reddened AGN classes, 
where the \hb\ measurement is impossible by definition.
The use of these lines coupled with a hard X--ray luminosity that is less affected by galaxy 
contamination and obscuration
\citep[which can be both correctly evaluated if $L_{\rm{X}}>10^{42}$ erg s$^{-1}$ and $N_{\rm{H}}<10^{24}$ cm$^{-2}$,][]{ranalli03,mineo14}, 
assures that these relations are able to reliably measure the BH mass also in AGN 
where the nuclear component is less prominent and/or contaminated by the hosting galaxy optical emission.
We can conclude that our new derived optical/NIR FWHM and hard X--ray luminosity
based virial relations can be of great help in measuring the BH mass in low-luminosity and absorbed AGN and therefore better measuring the complete (AGN1+AGN2) SMBH mass function. In this respect, in the future, a similar technique could
also be applied at larger redshift. For example, at redshift $\sim$2-3 the \pb\ line could be observed in the 1-5 $\mu$m wavelength range with NIRSPEC on the JamesWebb Space Telescope. 
While, after a straightforward recalibration, the rest-frame 14-195 keV X-ray luminosity could be substituted by the 10-40 keV hard X-ray band (which is as well not so much affected by obscuration for mildly absorbed, Compton Thin,
AGN). At redshift $\sim$2-3, in the observed frame, the 10-40 keV hard band roughly corresponds to the 2-10 keV energy range which is typically observed with the Chandra and XMM-Newton telescopes.

\begin{acknowledgements}
      We thank A. Graham whose useful comments improved the quality of the manuscript.
      Part of this work was supported by PRIN/MIUR 2010NHBSBE and PRIN/INAF 2014\_3.
\end{acknowledgements}

%
   \bibliographystyle{aa} 
   \bibliography{mybib} 
%

\end{document}